\documentclass[twocolumn,pra,aps]{revtex4}
\usepackage{graphicx}

\def\duzomniejsze{<\kern-.7mm<}
\def\duzowieksze{>\kern-.7mm>}

\def\textbf#1{{\bf #1}}
\def\beq{\begin{equation}}
\def\eeq{\end{equation}}
\def\be{\begin{equation}}
\def\ee{\end{equation}}
\def\ben{\begin{eqnarray}}
\def\een{\end{eqnarray}}
\def\beqa{\begin{eqnarray}}
\def\eeqa{\end{eqnarray}}
\def\eea{\end{array}}
\def\bea{\begin{array}}
\newcommand{\bei}{\begin{itemize}}
\newcommand{\eei}{\end{itemize}}
\newcommand{\bee}{\begin{enumerate}}
\newcommand{\eee}{\end{enumerate}}

\def\>{\rangle}
\def\<{\langle}

\begin{document}

\title{Mean parity of a single quantum excitation of some optical fields in thermal environments}

\author{Shang-Bin Li}

\affiliation{Shanghai research center of Amertron-global,
Zhangjiang High-Tech Park, \\
299 Lane, Bisheng Road, No. 3, Suite 202, Shanghai, 201204, P.R.
China}

\begin{abstract}
{\normalsize The mean parity (the Wigner function at the origin) of
excited binomial states, excited coherent states and excited thermal
states in thermal environments is investigated in details. It is
found that the single-photon excited binomial state and the
single-photon excited coherent state exhibit certain similarity in
the aspect of their mean parity in the thermal environment. We show
the negative mean parity can be regarded as an indicator of
nonclassicality of single-photon excitation of optical fields with a
little coherence, especially for the single-photon excited thermal
states.

}

\end{abstract}

\maketitle

Among various kinds of indicators of nonclassicality of optical
fields \cite{Kimble,Short,Dodonov,Hillery}, the partial negativity
of the Wigner function (PNWF) indicates the highly nonclassical
character of the optical fields. Quantum excitation of general
optical fields can always exhibit PNWF which will be destroyed and
eventually completely disappear at the same decay time
$\gamma{t}_c=\ln\frac{2+2n}{1+2n}$ in a thermal environment
\cite{Li20081}. The Wigner function $W(q,p)$ is proportional to the
expectation value of the parity operator that performs reflections
about the phase-space point ($q$, $p$) \cite{Royer77}. For an
optical field in the state $\rho$, the value of Wigner function at
the origin of the state can be derived by \cite{Cessa,Englert} \be
W(0,0)=\frac{2}{\pi}{\mathrm{Tr}}[(\hat{O}_e-\hat{O}_o)\hat{\rho}],\ee
where $\hat{O}_e\equiv\sum^{\infty}_{k=0}|2k\rangle\langle2k|$ and
$\hat{O}_o\equiv\sum^{\infty}_{k=0}|2k+1\rangle\langle2k+1|$ are the
even and odd parity operators respectively. Therefore, the value of
Wigner function at the origin equals the constant $\frac{2}{\pi}$
times the expectation value of parity operator for the state $\rho$.
Hereafter the expectation value of parity operator is named as "mean
parity".

There have been several kinds of experimental schemes for
reconstructing or measuring the Wigner distribution of the optical
fields or the motional states of trapped atoms
\cite{Smithey,Lutterbach}. Here, we briefly outline an operational
approach for measuring the Wigner function which is based on the
Fresnel transformation of measured Rabi oscillations
\cite{Lougovski2003}. Assuming a two-level atom (qubit) initially in
the excited state $|e\rangle$ is resonantly coupled with the field
prepared in the quantum state of interest whose dynamical behavior
is governed by the Jaynes-Cummings model in the interaction picture
$\lambda(a^{\dagger}\sigma_-+a\sigma_+)$. Via observing Rabi
oscillations of the two-level atom, the mean parity of the initial
state of interest can be derived as follows: \beqa
\frac{\pi}{2}W(0,0)&=&\frac{4}{\sqrt{i}}\int^{\infty}_{0}e^{i\tau^2/\pi}[P_g(\tau)-\frac{1}{2}]d\tau\nonumber\\
&&=\frac{2}{\sqrt{i}}\int^{\infty}_{0}e^{i\tau^2/\pi}[P_g(\tau)-P_e(\tau)]d\tau,\eeqa
where $P_g(\tau)$ ($P_e(\tau)$) is the probability of finding the
atom in the ground state $|g\rangle$ (the excited state $|e\rangle$)
as a function of dimensionless interaction time $\tau=\lambda{t}$.
Obviously, $P_g(\tau)-P_e(\tau)$ can be regarded as the mean parity
of the two-level atom. Thus, the initial mean parity of an optical
field is directly related to the Fresnel integral of the mean parity
of a resonant coupled two-level atom. Though the mean parity of a
quantum field contains less information than its full Wigner
function, it can partly determine the dynamical behaviors of Rabi
oscillation of the resonant coupled qubit, which is important in
those quantum information processes containing the resonant coupling
between nonclassical optical fields and qubit. Therefore, it is
desirable to investigate the dynamical behaviors of mean parity of
some kinds of nonclassical states in the thermal environments.

When the state $\rho$ evolves in the thermal environment, its
evolution can be described by \cite{Gardiner} \beqa
\frac{d\rho}{dt}&=&\frac{\gamma(n+1)}{2}(2a\rho{a}^{\dagger}-a^{\dagger}a\rho-\rho{a}^{\dagger}a)\nonumber\\
&&+\frac{\gamma{n}}{2}(2a^{\dagger}\rho{a}-aa^{\dagger}\rho-\rho{a}{a}^{\dagger}),
\eeqa where $\gamma$ represents dissipative coefficient and $n$
denotes the mean thermal photon number of the thermal environment.
In the thermal environment described by the master equation (2), the
time evolution Wigner function satisfies the following Fokker-Planck
equation \cite{Carmichael} \beqa
\frac{\partial}{\partial{t}}W(q,p,t)&=&\frac{\gamma}{2}(\frac{\partial}{\partial{q}}q+\frac{\partial}{\partial{p}}p)W(q,p,t)\nonumber\\
&&+\frac{\gamma(2n+1)}{8}(\frac{\partial^2}{\partial{q}^2}+\frac{\partial^2}{\partial{p}^2})W(q,p,t).\eeqa

Firstly, we will investigate the influence of the thermal noise on
the mean parity of the single-photon excited coherent state (ECS)
described by unnormalized wave vector $a^{\dagger}|\alpha\rangle$
\cite{Agarwal}. Let us recall the results in
Ref.\cite{Li20071,Li2007}, and the time-evolving Wigner function at
the origin of single-photon ECS in the thermal environment can be
derived as \beqa
W^S(0,0,\gamma{t})&=&\frac{2e^{\gamma{t}}[|\alpha|^2(1-c^2)^2+c^4-1]\exp[-\frac{2|\alpha|^2}{1+c^2}]}{\pi(1+|\alpha|^2)(1+c^2)^3},\nonumber\\
c&=&[(\exp(\gamma{t})-1)(1+2n)]^{1/2}.\eeqa
It is easy to check that
\beqa W^S(0,0,\gamma{t})<0~ {\mathrm{For}}~ \gamma{t}<\gamma{t}_c ~{\mathrm{and}} ~|\alpha|<1\nonumber\\
W^S(0,0,\gamma{t})\geq0 ~{\mathrm{For}} ~\gamma{t}\leq\gamma{t}_{c1}~ {\mathrm{and}}~ |\alpha|\geq1\nonumber\\
W^S(0,0,\gamma{t})<0~ {\mathrm{For}}~
\gamma{t_{c1}}<\gamma{t}<\gamma{t}_{c}~ {\mathrm{and}}~
|\alpha|\geq1, \eeqa where \be \gamma{t}_c=\ln\frac{2+2n}{1+2n}, \ee
and \be
\gamma{t}_{c1}=\ln[\frac{2|\alpha|^2(1+n)+2n}{(1+2n)(1+|\alpha|^2)}]\ee
The above results imply the mean parity of the initial pure single
photon ECS exhibits transition-like behavior with a critical point
at $|\alpha|=1$. When $|\alpha|<1$, the pure single photon ECS has
negative mean parity. Otherwise, it has positive definite mean
parity. The negativity of mean parity of the single photon ECS with
$|\alpha|<1$ will be maintained until the threshold decay time
$\gamma{t}_c$ beyond which the Wigner function of the thermal
dissipative single photon ECS become positive definite over the
whole phase space. The mean parity of the single-photon ECS with
$|\alpha|>1$ is positive before the threshold decay time
$\gamma{t}_{c1}$, and then becomes negative until $\gamma{t_c}$. In
this sense, the mean parity may be suitable to be regarded as one of
the indicators of the quantum-classical transition of the
single-photon ECS.

In what follows, excited binomial states (EBS) are chosen as the
example to investigate their dynamical behaviors of Wigner
distribution and mean parity in the thermal environment. The EBSs
have been introduced \cite{Wang2000}, which represent the
intermediate non-Gaussian state between quantum Fock state and ECS.
The EBS of the radiation field can be generated by repeated
application of the photon creation operator on binomial states
\cite{Wang2000}. The binomial states of optical fields can be
generated in some nonlinear processes \cite{Lee,Dattoli,Fu}. Then,
by a scheme similar to that preparing the ECS \cite{Agarwal}, in
which the excited atoms pass through a cavity and provide the field
in the cavity is initially in a binomial state, one can produce the
EBS. If a traveling optical field in the binomial state has been
produced, one can also adopt the experimental scheme of Zavatta et
al. \cite{Zavatta2004} to generate the single-photon-excited
binomial state.

\begin{figure}
\centerline{\includegraphics[width=6.3cm]{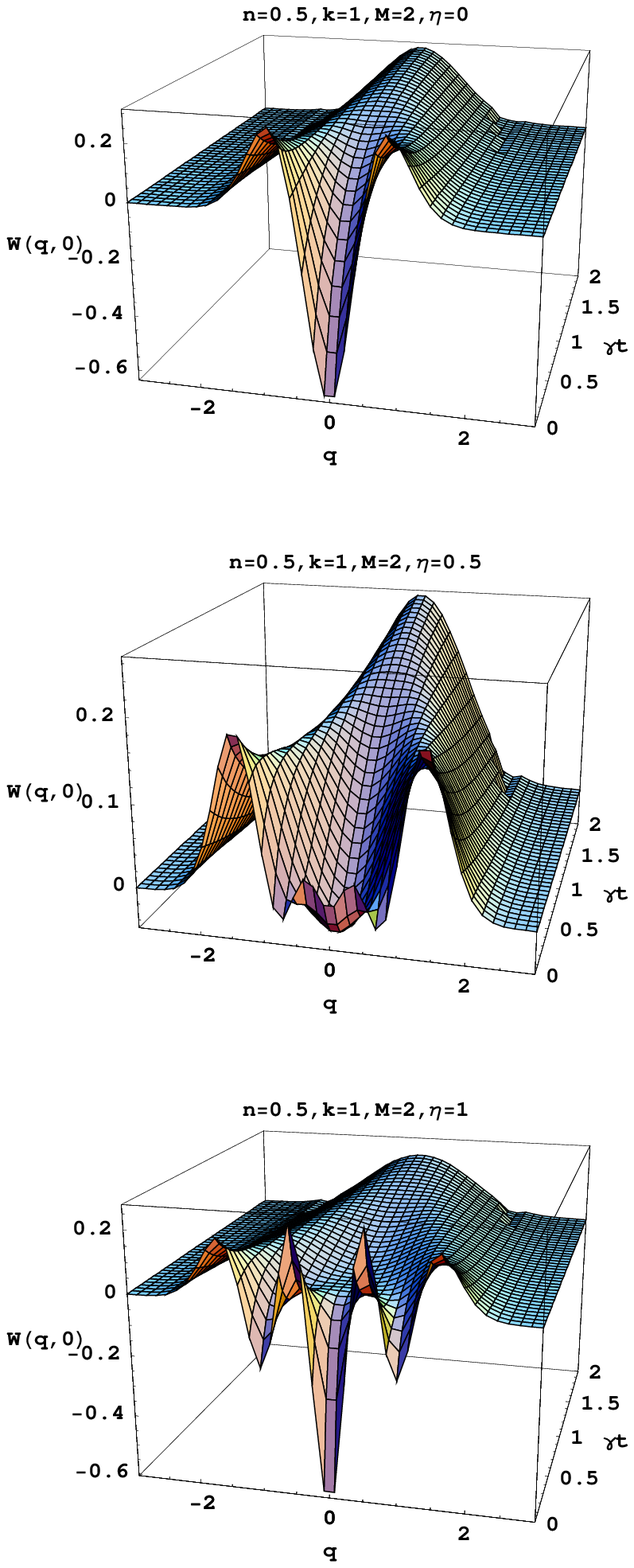}}
\caption{The cross section $W(q,0)$ of Wigner functions of the EBSs
$|1,0,2\rangle\equiv|1\rangle$, $|1,0.5,2\rangle$, and
$|1,1,2\rangle\equiv|3\rangle$ in the thermal environment with
$n=0.5$ are plotted as the function of decay time $\gamma{t}$.}
\end{figure}
Let us briefly recall the definition of the EBS \cite{Wang2000}. The
EBS is defined by \be
|k,\eta,M\rangle=N(k,\eta,M)a^{\dagger{k}}|\eta,M\rangle, \ee, where
\be
|\eta,M\rangle=\sum^M_{l=0}(C^l_M)^{1/2}\eta^l(1-\eta^2)^{(M-l)/2}|l\rangle
\ee is the binomial state \cite{Stoler}. Here $k$, $l$ and $M$ are
integers. $C^l_M\equiv\frac{M!}{l!(M-l)!}$ is the binomial
coefficient, $\eta$ is real number with $0\leq{\eta}\leq1$.
$|l\rangle$ is Fock state. $a^{\dagger}$ ($a$) is the creation
operator (annihilation operator) of the optical mode. $N(k,\eta,M)$
is normalization constant of EBS, which is given by \be
N(k,\eta,M)=[\frac{\eta^{2M}(M+k)!}{M!}{}_{2}F_1(-M,-M;-M-k;\frac{\eta^2-1}{\eta^2})]^{-\frac{1}{2}}\ee,
where ${}_2F_1(\varepsilon,\zeta;\kappa;x)$ is the hypergeometric
function. The statistics properties of pure EBSs have been
investigated in the past few years \cite{Wang2000}. In
Ref.\cite{Franco}, a scheme for preparing two photon binomial state
in single mode high-Q cavity has been proposed. In fact, by slightly
changing that scheme, one can generate the EBS with $k=1$, and
$M=2$. Let an excited atom pass through the cavity initially in the
two photon binomial state, one can produce the EBS with $k=1$, and
$M=2$. In Ref.\cite{Darwish}, the analytical expressions of Wigner
function of the EBSs have been derived. Substituting them into
Eq.(4) as the initial conditions and numerically solving this
partial differential equation, one can obtain the time evolving
Wigner function. As an illustration, in Fig.1, we have plotted the
evolution of the cross-section of Wigner distribution of
$|1,\eta,2\rangle$ in the thermal environment labeled with $n=0.5$
for three different values of $\eta$. The partial negativity of the
cross section of Wigner function indicates the nonclassical nature
of the single-photon EBS. The thermal noise destroys the partial
negativity of the Wigner function. This figure explicitly exhibits
the role of the single-photon-occurring probability $\eta^2$ in its
corresponding Wigner distribution.

\begin{figure}
\centerline{\includegraphics[width=6.3cm]{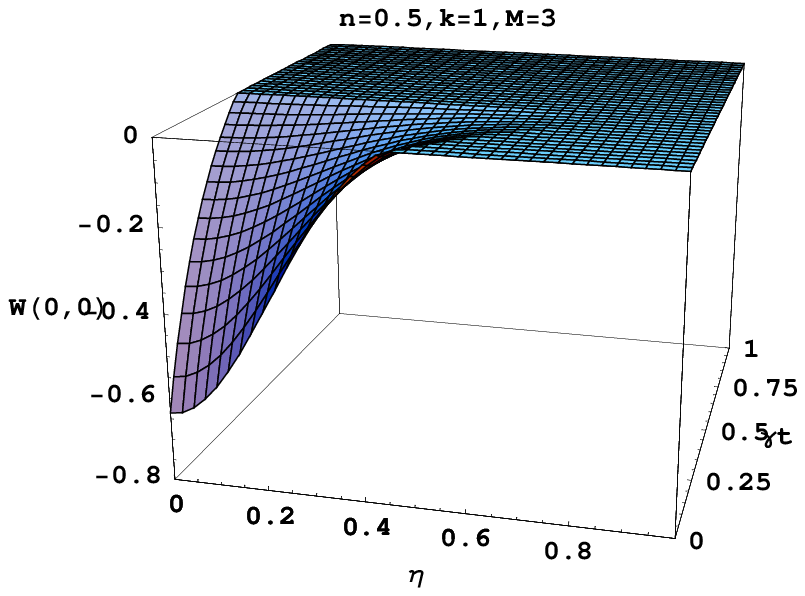}}
\caption{The value $W(0,0)$ of Wigner function at the origin of the
EBS with $k=1$, and $M=3$ in the thermal environment with $n=0.5$ is
plotted as the function of decay time $\gamma{t}$ and $\eta$.}
\end{figure}
\begin{figure}
\centerline{\includegraphics[width=6.3cm]{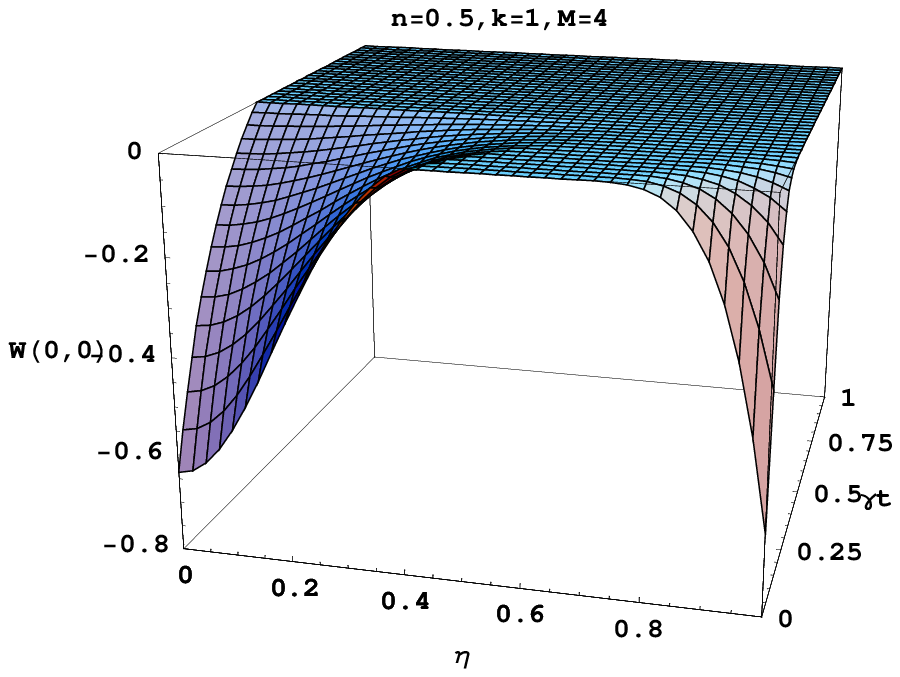}}
\caption{The value $W(0,0)$ of Wigner function at the origin of the
EBS with $k=1$, and $M=4$ in the thermal environment with $n=0.5$ is
plotted as the function of decay time $\gamma{t}$ and $\eta$.}
\end{figure}
\begin{figure}
\centerline{\includegraphics[width=6.3cm]{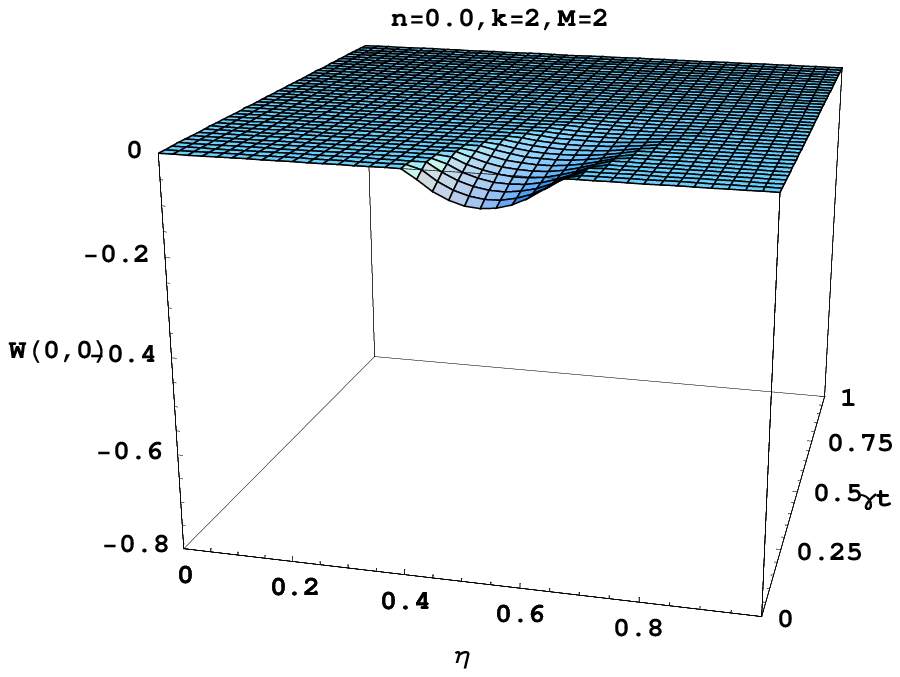}}
\caption{The value $W(0,0)$ of Wigner function at the origin of the
EBS with $k=2$, and $M=2$ in the photon loss channel with $n=0$ is
plotted as the function of decay time $\gamma{t}$ and $\eta$.}
\end{figure}

In Fig.2, $W(0,0)$ of Wigner function at the origin of the
single-photon EBS with $M=3$ in the thermal environment with $n=0.5$
is plotted as the function of decay time $\gamma{t}$ and $\eta$.
With varying of $\eta$ from 0 to 1, $W(0,0)$ exhibits different
dynamical behaviors in the thermal environment: For small values of
$\eta<\eta^{(3)}_1\simeq0.4$, $W(0,0)$ is always negative before the
threshold decay time $\gamma{t}_c$. For $\eta^{(3)}_1\leq\eta<1$,
$W(0,0)$ is initially non-negative and then becomes negative until
the threshold decay time $\gamma{t}_c$. As $\eta=1$, $W(0,0)$ is
always non-negative. In Fig.3, $W(0,0)$ of Wigner function at the
origin of the single-photon EBS with $M=4$ in the thermal
environment with $n=0.5$ is plotted as the function of decay time
$\gamma{t}$ and $\eta$. In this case, for small values of
$\eta<\eta^{(4)}_1\simeq0.4$, $W(0,0)$ is always negative before the
threshold decay time $\gamma{t}_c$. For
$\eta^{(4)}_1\leq\eta<\eta^{(4)}_2\simeq\sqrt{2}/2$, $W(0,0)$ is
initially non-negative, then becomes negative until the threshold
decay time $\gamma{t}_c$. For $\eta^{(4)}_2<\eta<1$, $W(0,0)$ is
initially negative, then becomes non-negative, and becomes negative
again until the threshold decay time $\gamma{t}_c$. We have also
calculated the cases with larger values of $M$, and found similar
behaviors. Now for single-photon EBSs, the dependence of
thermal-noise-induced behavior of $W(0,0)$ on $\eta$ can be
classified: For the cases with $M=2m+1, (m=1,2,...)$,
$0\leq\eta<\eta^{(M)}_1$, $W(0,0)$ is negative before the threshold
decay time $\gamma{t}_c$. For the cases with $M=2m+1, (m=1,2,...)$,
$\eta^{(M)}_1\leq\eta<1$, $W(0,0)$ is initially positive and then
become negative until the threshold decay time $\gamma{t}_c$. The
critical point $\eta^{(M)}_1$ decreases with the increase of $M$.
For the cases with $M=2m, (m=1,2,...)$, $0\leq\eta<\eta^{(M)}_1$,
$W(0,0)$ is negative before the threshold decay time $\gamma{t}_c$.
For the cases with $M=2m, (m=1,2,...)$,
$\eta^{(M)}_1\leq\eta<\eta^{(M)}_2$, $W(0,0)$ is initially positive
and then become negative until the threshold decay time
$\gamma{t}_c$. For the cases with $M=2m, (m=1,2,...)$,
$\eta^{(M)}_2\leq\eta<1$, $W(0,0)$ is initially negative and then
become positive and become negative again until the threshold decay
time $\gamma{t}_c$. $\eta^{(M)}_2$ increases with $M$.

The inherent physics picture may be explained as follows: When
$\eta$ equals zero or one, the EBS reduce to the Fock state. The
parity property of Fock states tell us that only those Fock states
with odd photon numbers have negative value of Wigner function at
the origin. For very small value of $\eta$, the single-photon EBS
looks like the single-photon Fock state. For intermediate value of
$\eta^{(M)}_1<\eta<\eta^{(M)}_2$, the single-photon EBS is similar
to the single-photon ECS with $|\alpha|>1$ in the aspect of thermal
noise induced evolution of mean parity.

Up to now, we have understood the dynamical characteristics of mean
parity of both single-photon ECS and EBS in the thermal environment.
These states have not population in vacuum state. We also
investigate the states with population in vacuum state such as the
binomial state in the thermal environment, and the results show they
more rapidly completely lose the PNWF than those states with zero
population in the vacuum state. In other words, the initial possible
negative mean parity of the binomial state is more fragile against
the thermal noise than the single-photon EBS. The negative mean
parity is a sufficient condition for nonclassicality of quantum
optical fields, thus single-photon EBS may have advantage in
potential applications in quantum optical information processes.

While for some two-photon EBSs with initial negative mean parity,
our numerical result indicates that the negative mean parity of
two-photon EBS are more fragile against the thermal noise than the
single-photon EBS. In Fig.4, $W(0,0)$ of the two-photon EBS with
$M=2$ in photon loss channel with $n=0$ is plotted as the function
of decay time $\gamma{t}$ and $\eta$. Initially, the mean parity is
negative only for those states with intermediate $\eta$. For small
value of $\eta$, the mean parity is always positive. For
intermediate value of $\eta$, the mean parity maintains its negative
value in a period shorter than $\gamma{t}_c$, and then perpetually
becomes positive.

For the single-photon excited thermal state \cite{Agarwal1992}, it
is also easy to obtain its time-evolving Wigner function at the
origin in the thermal environment as follows \cite{Li20081}: \beqa
W^T(0,0,\gamma{t})&=&\frac{\kappa}{\pi{\xi^3}}e^{\gamma{t}},\nonumber\\
\xi&=&2(\bar{n}-n)+(1+2n)e^{\gamma{t}},\nonumber\\
\kappa&=&-8(\bar{n}-n)(1+n)+2(1+2n)^2e^{2\gamma{t}}\nonumber\\
&&+4[\bar{n}(1+2n)-(1+2n)^2]e^{\gamma{t}}, \eeqa where $\bar{n}$
denotes the mean photon number of the initial thermal state. Its
mean parity is always negative before the threshold decay time
$\gamma{t}_c=\ln\frac{2+2n}{1+2n}$. In this case, the negative mean
parity can be regarded as a suitable indicator of nonclassicality of
single-photon excited thermal state in the thermal environment for
its monotonic corresponding to the volume of negative part of Wigner
function.

The above several examples do show that the negativity of mean
parity, as defined, is a useful indicator of the nonclassicality of
some specific class of non-classical fields. Especially, it exhibit
the advantage if used as an indicator of convenient experimental
analysis whether the thermal noise has completely destroyed the
negativity of Wigner distribution of any photon-added optical fields
or not. The results in Ref.\cite{Li20081} have demonstrated
arbitrary photon added optical fields completely lose their
negativity of Wigner distribution at the same threshold decay time
in the thermal environment labeled by the same effective
temperature, and their mean parity is zero at the threshold decay
time.

In summary, we have investigated the mean parity of single quantum
excitations of coherent states, binomial states and thermal states
in the thermal environment. The single quantum excitation usually
strengthen the negative mean parity. When the single quantum
excitation of classical or nonclassical optical fields lacks for
enough coherence, such as the single-photon excited thermal state,
ECS and EBS which have high overlap with single photon Fock state,
the initial negative mean parity monotonically increase and become
positive after the decay time $\gamma{t}_c$. In these cases, the
negative mean parity can be regarded as a good indicator of
nonclassicality of optical fields. While for those single-photon ECS
or EBS with sufficient large coherence, their mean parity will
exhibit transient negative even if initially the mean parity is
positive.

For single-photon ECS, its mean parity can be also regarded as the indicator for the transition from Fock-like state to
coherent-like state.

To experimentally investigate whether the thermal noise has
completely destroyed the PNWF,
one has to reconstruct the whole phase space Wigner distribution of
real-time states, though reconstruction
of the whole phase space Wigner distribution is still very complicated.
Thus, for some specific single-photon excitation of optical fields with a little coherence, to measure the mean parity may be a feasible method.




\end{document}